\documentclass[aps,prb,twocolumn,showpacs]{revtex4}
\pdfoutput=1
\usepackage{amsmath}
\usepackage{amssymb}
\usepackage{graphicx}
\usepackage{multirow}
\usepackage{color}

\newcommand{\forget}[1]{}
\newcommand{\bd}[1]{\boldsymbol{#1}}

\begin{document}

\title{Electron-solid and electron-liquid phases in graphene}

\author{M.E. Knoester$^1$}
\author{Z. Papi\'{c}$^2$}
\author{C. Morais Smith$^1$}
\affiliation{$^1$Institute for Theoretical Physics, Center for Extreme Matter and Emergent Phenomena, Utrecht
University, Leuvenlaan 4, 3584 CE Utrecht, The Netherlands}
\affiliation{$^2$School of Physics and Astronomy, University of Leeds, Leeds, LS2 9JT, United Kingdom}

\date{\today}

\begin{abstract}
We investigate the competition between electron-solid and quantum-liquid phases in graphene, which arise in partially filled Landau levels. The differences in the wave function describing the electrons in the presence of a perpendicular magnetic field in graphene with respect to the conventional semiconductors, such as GaAs, can be captured in a form factor which carries the Landau level index. This leads to a quantitative difference in the electron-solid and -liquid energies. For the lowest Landau level, there is no difference in the wave function of relativistic and non-relativistic systems. We compute the cohesive energy of the solid phase analytically using a Hartree-Fock Hamiltonian. The liquid energies are computed analytically as well as numerically, using exact diagonalization. We find that the liquid phase dominates in the $n=1$ Landau level, whereas the Wigner crystal and electron-bubble phases become more prominent in the $n=2$ and $n=3$ Landau level.
\end{abstract}

\pacs{73.20.Qt, 73.22.Pr, 73.43.Nq}

\maketitle

\section{Introduction}
Since it experimental realization in 2005 \cite{Geim}, graphene has attracted much attention in the scientific world, not only for its potential use in technological devices, but also because of its linear energy dispersion, which allows for the realization of relativistic electrons in table-top experiments \cite{reviewCastroNeto}. Due to the presence of Dirac cones in the energy spectrum, the integer quantum Hall effect (IQHE) exhibits anomalous features, as predicted theoretically \cite{gusynin} and observed experimentally \cite{geimqhe}. In addition, the fractional quantum Hall effect (FQHE), arising due to electronic interactions, was also observed to display some anomalous behavior in graphene~\cite{fqhe1,fqhe2,fqhe3,fqhe4,fqhe5,fqhe6,fqhe7}.

In conventional two-dimensional electron systems (2DES) like GaAs, strong magnetic field restricts the dynamics of electrons to a singe Landau level. Because of the reduced Hilbert space, the form of the effective Coulomb potential felt by electrons can then vary from one Landau level to the next. Typically, the lowest two partially filled Landau levels give rise to a plethora of FQHE liquid phases, such as the Laughlin \cite{laughlin} and ``composite-fermion" states \cite{jain89}. When the electron filling factor becomes very low, electrons instead form a Wigner crystal \cite{lamgirvin}. Additionally, in higher Landau levels an effective short-range attractive interaction favors the formation of electron stripes or electron-bubble solids with two, three or more electrons per site \cite{sb1, sb2, sb3, sb4, sb5, sb6, sb7, sb8, sb9, sb10}. A competition between the Wigner crystal (WC), electronic-bubble, and the quantum-liquid phase gives rise to a reentrant IQHE 
~\cite{goerbig,riqhe1,riqhe2,riqhe3,riqhe4,riqhe5,riqhe6}. This effect consists of a series of first-order quantum phase transitions, much like the classical phase transition from ice to water, which occurs upon increasing the temperature. In GaAs, the quantum phase transition can repeat itself upon increasing the magnetic field, producing an electron-solid that melts into a normal liquid, which becomes an incompressible quantum-liquid at particular values of the filling factor, and then solidifies again upon increasing the magnetic field even further. The phenomenon is accompanied by reentrant plateaus in the Hall resistance, which become quantized at the value of the resistivity of the most nearby integer, every time when the electrons in the partially filled Landau level solidify.  The question then arises whether this effect can be also observed in graphene. As in the conventional 2DES, the Landau level wave functions for electrons in graphene can be written as a generic form factor multiplied by a Gaussian~\cite{fngraphene, fngraphene2}. However, except for the lowest Landau level, which exhibits the same behavior as in the 2DES, in graphene this form factor mixes two adjacent Landau levels, and the resulting effective Coulomb potential is generally different from GaAs~\cite{goerbigreview}. 

In this paper, we investigate how the theory of the reentrant IQHE in conventional 2DES should be modified, in order to describe the competition between electron-solid and electron-liquid phases in graphene. We find that upon an appropriate rescaling that takes into account the special form factor of graphene, a universal scaling behavior emerges for the real-space interaction. As for GaAs \cite{goerbig, epl}, the effective interaction contains a short-range attractive part, which leads to the formation of electron-bubble phases.  First, we calculate the cohesive energies of the WC and of the electron-bubble phase using the Hartree-Fock approximation. Subsequently, we calculate the energy of the Laughlin liquid analytically and the exact ground-state energies of finite systems by numerical diagonalization. We find that in the $n=1$ Landau level, the Laughlin liquid has much lower energy than the WC and it is improbable that electron-solid phases would be observed, contrarily to GaAs. In the $n=2$ and $n=3$ Landau levels, we find these phases should be observable for certain ranges of the filling factor. We have also considered the effect of impurities in the sample by modeling them by a Gaussian impurity potential. It turns out that the electron-solid energies can be lowered by the impurities, especially at low fillings, thereby washing out some of the FQH states.

The outline of this paper is as follows: in Sec.~\ref{sec:model}, we explain the model for electronic interactions in graphene in partially filled Landau levels and the universal scaling behavior. Subsequently, we derive the energies of the electron-solid and electron-liquid phases in Sec.~\ref{sec:solid} and~\ref{sec:liquid}, respectively. The phase diagrams arising from the competition between the phases are constructed in Sec.~\ref{sec:competition} for several partially filled Landau levels, and our conclusions are provided in Sec.~\ref{sec:conclusion}.
Appendix \ref{appED} contains details about the numerical calculations using the exact-diagonalization method.

\section{The model}\label{sec:model}
We consider spinless electrons and restrict their dynamics to the $n$-th Landau level. In the absence of inter-Landau-level excitations, we can express the partial filling of the $n$-th Landau level as $\bar{\nu}=\nu-[\nu]=N/N_\Phi$, where $N$ is the number of electrons in the topmost Landau level and $N_\Phi=A/2\pi \ell_B^2$ is the degeneracy in each Landau level, given in terms of the area of the sample $A$ and the magnetic length $\ell_B=\sqrt{\hbar/eB}$. The wave functions in the $n$-th Landau level are given by $\psi_\sigma(\bd{r})=e^{i\sigma\bd{K}\cdot\bd{r}}\chi_\sigma(\bd{r})$, for $\sigma=\pm 1$, with $\pm\bd{K}$ the corners of the Brillouin zone (the K and K$^\prime$ valleys) and~\cite{fngraphene}
\begin{align*}
\chi_+(\bd{r})=\frac{1}{\sqrt{2}}\sum_{n,m}
\begin{pmatrix}
i\sqrt{1+\delta_{n,0}}\,\langle\bd{r}| |n|,m\rangle\\
\text{sgn}(n)\langle\bd{r}| |n|-1,m\rangle
\end{pmatrix},
\\
\chi_-(\bd{r})=\frac{1}{\sqrt{2}}\sum_{n,m}
\begin{pmatrix}
\text{sgn}(n)\langle\bd{r}| |n|-1,m\rangle\\
i\sqrt{1+\delta_{n,0}}\,\langle\bd{r}| |n|,m\rangle
\end{pmatrix}.
\end{align*}
The density operator in the $n$-th Landau level is then the sum of the two sublattice density operators $\rho_\alpha(\bd{r})=\sum_{\sigma,\sigma'}\psi^{\dagger}_{\alpha,\sigma}(\bd{r})\psi_{\alpha,\sigma'}(\bd{r})$. In reciprocal space it is given by
\begin{align}\label{eq:rho}
\rho^n(\bd{q})=\rho_1^n(\bd{q})+\rho_2^n(\bd{q})=\sum_{\sigma,\sigma'}F_n^{\sigma,\sigma'}(\bd{q})\bar{\rho}^{\sigma,\sigma'}(\bd{q}),
\end{align}
where the projected density operators $\bar{\rho}^{\sigma,\sigma'}(\bd{q})=\sum_{m,m'}\langle m| e^{-i[\bd{q}+(\sigma-\sigma')\bd{K}]\cdot\bd{R}}|m'\rangle c^{\dagger}_{n,m,\sigma}c_{n,m',\sigma'}$ project onto the lowest Landau level~\cite{fngraphene}. Here, $c^{\dagger}_{n,m,\sigma}$ ($c_{n,m,\sigma}$) creates (annihilates) a state $|n,m\rangle$ in the $n$-th Landau level in the valley $\sigma$ and $\bd{R}$ is the guiding center operator. The components of the form factor, which capture the Landau-level dependence, are given in terms of Laguerre polynomials by
\begin{align*}
F_n^{\sigma,\sigma}=&\frac{1}{2}\left[L_{|n|}\left(\frac{q^2\ell_B^2}{2}\right)+L_{|n|-1}\left(\frac{q^2\ell_B^2}{2}\right)\right]e^{-q^2\ell_B^2/4},\\
F_n^{\sigma,-\sigma}=&\frac{\lambda i\ell_B[q+q^*-\sigma(K+K^*)]}{2\sqrt{2|n|}}\\
&\times L_{|n|-1}^1\left(\frac{\ell_B^2|\bd{q}-\sigma\bd{K}|^2}{2}\right)e^{-|\bd{q}-\sigma\bd{K}|^2\ell_B^2/4},
\end{align*}
where we used the complex notation $q=q_x-iq_y$ and $K=K_x-iK_y$~\cite{fngraphene}.

Since all electrons under consideration are in the same Landau level, the kinetic energy is quenched and the Hamiltonian consists only of the Coulomb interaction $v(\bd{r})$ between the electrons,
\begin{align*}
H=\frac{1}{2}\int d^2r d^2r' \rho_n(\bd{r}) v(\bd{r}-\bd{r}') \rho_n(\bd{r}').
\end{align*}
Transforming it to reciprocal space and using the projected density operators defined in Eq.~\eqref{eq:rho}, we can write this Hamiltonian as~\cite{epl,fngraphene}
\begin{align*}
H=\frac{1}{2}\sum_{\sigma_1,\cdots,\sigma_4}\sum_{\bd{q}}v_n^{\sigma_1,\cdots,\sigma_4}(\bd{q})\bar{\rho}^{\sigma_1\sigma_3}(-\bd{q})\bar{\rho}^{\sigma_2\sigma_4}(\bd{q}),
\end{align*}
where we absorbed the form factor into an effective interaction, which is given by~\cite{fngraphene, fngraphene2}
\begin{align*}
v_n^{\sigma_1,\cdots,\sigma_4}(\bd{q})=v(q)F_n^{\sigma_1\sigma_3}(-\bd{q})F_n^{\sigma_2\sigma_4}(\bd{q}),
\end{align*}
with $v(q)=2\pi e^2/\epsilon q$ the usual Coulomb interaction in reciprocal space, where $\epsilon$ denotes the dielectric constant. Since terms of the form $F_n^{\sigma,\sigma}(\mp\bd{q})F_n^{\sigma',-\sigma'}(\pm\bd{q})$ or $F_n^{\sigma,-\sigma}(-\bd{q})F_n^{\sigma,-\sigma}(\bd{q})$ are exponentially small in $a/\ell_B$ and backscattering terms of the form $F_n^{\sigma,-\sigma}(-\bd{q})F_n^{-\sigma,\sigma}(\bd{q})$ are algebraically small in $a/\ell_B$~\cite{fngraphene}, we may write the interaction Hamiltonian up to leading order in perturbation theory as
\begin{align}\label{eq:ham}
H=\frac{1}{2}\sum_{\sigma,\sigma'}\sum_{\bd{q}}v_n^g(q)\bar{\rho}_\sigma(-\bd{q})\bar{\rho}_{\sigma'}(\bd{q}),
\end{align}
where we defined $\bar{\rho}_\sigma(\bd{q})\equiv\bar{\rho}^{\sigma\sigma}(\bd{q})$ and $v_n^g(q)$ is the effective interaction in graphene, given by
\begin{equation}
v_n^g(q)=v(q)[F_n^g(q)]^2, \label{eq:vq}
\end{equation}
with $F_n^g(q)$ the graphene form factor
\begin{eqnarray}\label{eq:graphenepp}
\nonumber F_{n\neq 0}^g(q) &=& \frac{1}{2}  \left[  L_{|n|}\left(\frac{q^2\ell_B^2}{2}\right)  \right. \\
&& \left. +L_{|n|-1}\left(\frac{q^2\ell_B^2}{2}\right)  \right]  e^{-q^2\ell_B^2/4},  \\
F_0^g(q) &=& e^{-q^2\ell_B^2/4}. 
\end{eqnarray}
Notice that for $n=0$, the form factor coincides with the one for conventional 2DEGs~\cite{goerbig} and for $n>0$ it averages two of them over two adjacent GaAs Landau levels.

\subsection{Effective interaction}

For GaAs, the effective Coulomb potential in Landau level $n$ is
\begin{align*}
v_n(q)=v(q) \left[ L_n\left( \frac{|q|^2\ell_B^2}{2}\right) \right]^2 e^{-|q|^2\ell_B^2/2}.
\end{align*}
As shown by Goerbig et al.~\cite{epl}, by rewriting the effective interaction 
in real space and subsequently scaling it by $R_C/\ell_B$, while simultaneously scaling the coordinates by $R_C$, a universal curve arises for all $n>0$, where $R_C$ denotes the cyclotron radius. In the scaled interaction, a shoulder emerges at a universal length scale of $2R_C$, which can lead to the formation of electron-bubble phases. 

\begin{figure}
\centering
\includegraphics[width=\columnwidth]{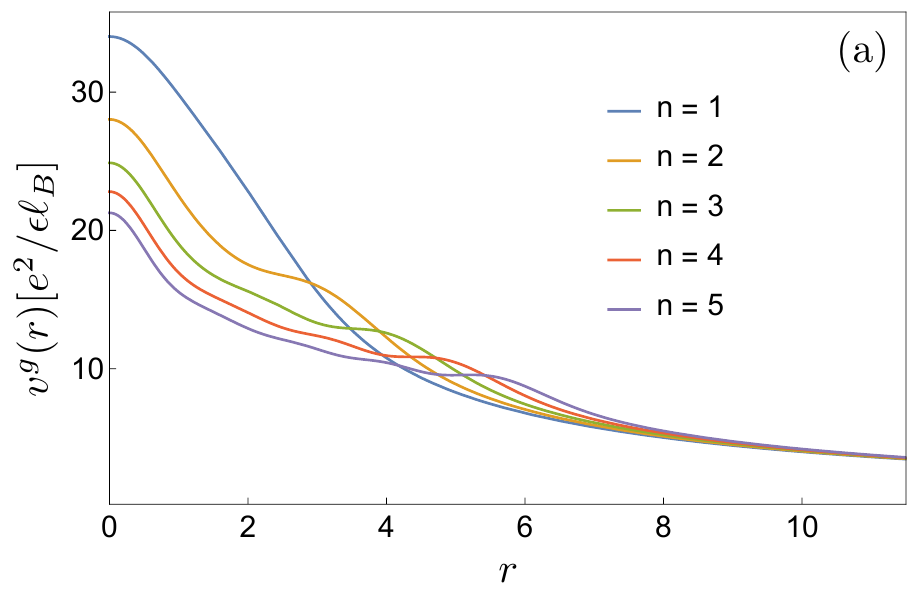}
\includegraphics[width=\columnwidth]{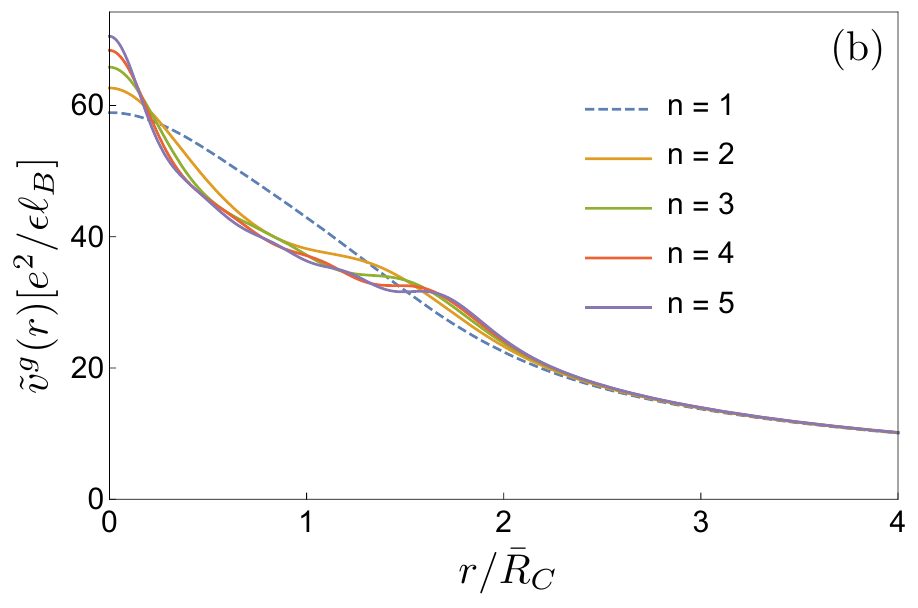}
\caption{(Color online) (a) Effective real-space potential in various Landau levels. (b) The rescaled potential. A universal length scale emerges, except in the $n=1$ Landau level.}
\label{fig:veff} 
\end{figure}

The real-space interaction for various Landau levels in graphene is shown in Fig.~\ref{fig:veff}a. To construct a scaled potential for graphene, we must take into account that the appropriate  cyclotron radius must also be averaged over two Landau levels, because of the averaging of two form factors. Hence, it reads 
\begin{align*}
\bar{R}_C&=\frac{\ell_B}{2}\left(\sqrt{2n+1}+\sqrt{2(n-1)+1}\right).
\end{align*} 
The scaled real-space potential $\tilde{v}_n(r)$, defined by
\begin{align*}
\frac{\tilde{v}^g_n(r/\bar{R}_C)}{\bar{R}_C/\ell_B}=v^g_n(r),
\end{align*}
is shown in Fig.~\ref{fig:veff}b. Here, we can see a universal length scale of $2\bar{R}_C$ emerging in the scaled potential, except in the $n=1$ Landau level, which remains scale free. This is a consequence of the averaging of the $n=1$ and $n=0$ Landau level, since the lowest Landau level interaction is also scale free~\cite{goerbig}. Hence, we expect  only a Wigner crystal or a liquid phase in the $n=1$ Landau level, since the existence of bubbles is intrinsically linked to the effective attractive interaction at length scales $\bar{R}_C<r<2\bar{R}_C$, which manifests as a shoulder (a plateau) in the effective rescaled potential for $n>1$.

\section{Solid phase}\label{sec:solid}
\begin{figure}
\centering
\includegraphics[width=\columnwidth]{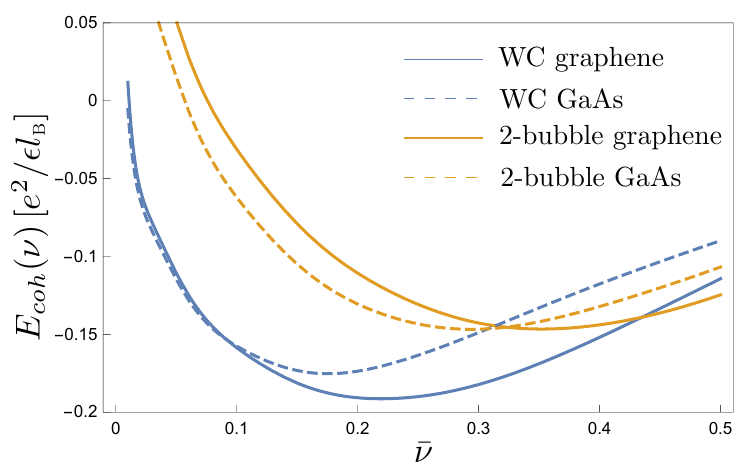}
\caption{(Color online) Cohesive energy of the WC and 2-bubble phases for conventional 2DEGs~\cite{goerbig} (dashed line) and graphene (solid line) in the $n=1$ Landau level.}
\label{fig:compared}
\end{figure}

As a first step in describing electron-solid phases (like Wigner crystals and bubble phases) in graphene, we consider a fully spin- and valley-polarized state, such that the interaction Hamiltonian reads
\begin{align*}
H=\frac{1}{2}\sum_{\bd{q}}v_n^g(q)\bar{\rho}(-\bd{q})\bar{\rho}(\bd{q}).
\end{align*}
As shown by Goerbig et al.~\cite{goerbig}, the electron-solid phases in GaAs are accurately described by the Hartree-Fock Hamiltonian 
\begin{align*}
H_{HF}=\frac{1}{2}\sum_{\bd{q}} u_n^{HF}(\bd{q})\langle \bar{\rho}(-\bd{q})\rangle \bar{\rho}(\bd{q}),
\end{align*}
where
\begin{align*}
u_n^{HF}(\bd{q})&=u_n^H(\bd{q})-u_n^F(\bd{q}),
\end{align*}
where the Hartree term is simply the effective interaction $u_n^H(\bd{q})=v_n(q)$ and the Fock exchange term is given in terms of $v_n(q)$ by
\begin{align*}
u_n^F(\bd{q})=\frac{1}{N_\Phi}\sum_{\bd{p}} v_n(p) e^{-i(p_yq_x-q_y p_x)\ell_B^2}.
\end{align*}
For graphene, we can use the same Hartree-Fock Hamiltonian by simply substituting the appropriate effective potential $v_n^g(q)$. The Hartree term is then simply $u_n^H(q)=v_n^g(q)$, and the exchange potentials $u_n^F(q)$ in graphene can be computed explicitly. In the first few Landau levels they read
\begin{eqnarray}
\nonumber u_1^F(q) &=& \frac{e^{-\tilde{q}^2/4}}{32n_\Phi}\sqrt{\frac{\pi}{2}}\left[(22+2\tilde{q}^2+\tilde{q}^4)I_0\left(\frac{\tilde{q}^2}{4}\right)\right.\\
&&  \left.-(4\tilde{q}^2+\tilde{q}^4)I_1\left(\frac{\tilde{q}^2}{4}\right)\right],
\end{eqnarray}
\begin{eqnarray}
\nonumber u_2^F(q) &=& \frac{e^{-\tilde{q}^2/4}}{512n_\Phi}\sqrt{\frac{\pi}{2}}\\ 
\nonumber && \times\left[(290 - 12 \tilde{q}^2 + 28 \tilde{q}^4 - 2 \tilde{q}^6 + \tilde{q}^8)I_0\left(\frac{\tilde{q}^2}{4}\right)\right.\\
&& \left.-(56\tilde{q}^2 + 30 \tilde{q}^4 + \tilde{q}^8)I_1\left(\frac{\tilde{q}^2}{4}\right)\right],
\end{eqnarray}
\begin{eqnarray}
\nonumber u_3^F(q)&=& \frac{e^{-\tilde{q}^2/4}}{18432n_\Phi}\sqrt{\frac{\pi}{2}}\left[(9270 - 1458 \tilde{q}^2 + 1809 \tilde{q}^4 \right.\\
\nonumber && \left. - 360 \tilde{q}^6 + 114 \tilde{q}^8 - 14 \tilde{q}^{10} + \tilde{q}^{12})I_0\left(\frac{\tilde{q}^2}{4}\right)\right.\\
\nonumber && \left.-(1836\tilde{q}^2 + 1563 \tilde{q}^4 - 192 \tilde{q}^6 + 92 \tilde{q}^8\right.\\
&& \left. - 12 \tilde{q}^{10} + \tilde{q}^{12})I_1\left(\frac{\tilde{q}^2}{4}\right)\right],
\end{eqnarray}
where $n_\Phi=N_\Phi/A$ and $\tilde{q}\equiv \ell_B q$. Using these potentials, the cohesive energy of an $M$-electron bubble phase can be written as~\cite{goerbig}
\begin{align*}
E_{coh}^B(n;M,\bar{\nu})&=\frac{n_\Phi\bar{\nu}}{M}\sum_l u_n^{HF}(\bd{G}_l)\frac{J_1(\sqrt{2M}\ell_B|\bd{G}_l|)^2}{\ell_B^2|\bd{G}_l|^2},
\end{align*}
where $\bd{G}_l$ are the lattice vectors of the reciprocal triangular lattice that is formed by the electrons. In Fig.~\ref{fig:compared}, we have compared the cohesive energy of the relevant solid phases in the $n=1$ Landau level for graphene and GaAs. The Wigner crystal phase clearly has lower energy in graphene than in GaAs for filling factors larger than 0.1, whereas the 2-electron-bubble phase in graphene has a higher energy for small filling factors and a lower energy for filling factors larger than 0.3, approximately.

When comparing our results to numerical calculations of the bubble energies using Green's functions, done by Zhang et al.~\cite{zhang}, we can conclude that the results are in qualitative agreement, as can be seen from Table~\ref{tab:zhang}. The transition points between the Wigner crystal and the 2-electron-bubble phase and the transition between the 2-electron-bubble and 3-electron-bubble phase occur at similar values of the filling factor. The only significant difference is in the $n=1$ Landau level. Furthermore, they also calculated the energy of the oblique Wigner crystal and it turns out that this phase becomes the ground state near half filling~\cite{zhang}. Hence, it does not interfere with the solid phases that we consider, and we neglect this possibility in the following.

\begin{table}
\centering
\begin{tabular}{p{0.05\textwidth}|p{0.17\textwidth}|p{0.12\textwidth}|p{0.10\textwidth}}
LL&Transtition point &Numerical results by Zhang et al.~\cite{zhang} & Our analytical results\\
\hline
$n=1$& WC $\rightarrow$ 2-bubble &$\bar{\nu}= 0.62$ & $\bar{\nu}= 0.44$\\
\hline
\multirow{2}{*}{$n=2$}& WC $\rightarrow$ 2-bubble & $\bar{\nu}= 0.28$ & $\bar{\nu}= 0.25$\\
& 2-bubble $\rightarrow$ 3-bubble & $\bar{\nu}= 0.43$ & $\bar{\nu}= 0.43$\\
\hline
\multirow{2}{*}{$n=3$}& WC $\rightarrow$ 2-bubble & $\bar{\nu}= 0.18$ & $\bar{\nu}= 0.18$\\
& 2-bubble $\rightarrow$ 3-bubble & $\bar{\nu}= 0.30$ & $\bar{\nu}= 0.30$
\end{tabular}
\caption{Comparison between our analytical results and numerical results by Zhang et al.~\cite{zhang} of the filling factors at which the transition of various phases occurs.}
\label{tab:zhang}
\end{table}

\section{Liquid phase}\label{sec:liquid}

In terms of FQHE liquid candidate states, we confine ourselves to the Laughlin states \cite{laughlin} at filling factors $\bar{\nu}_L=1/(2s+1)$, with $s$ integer.   The total energy of such a state is then given by~\cite{goerbig}
\begin{align}\label{eq:utotal}
U=E^L_{coh}(n,s)-\frac{\bar\nu}{2A}\sum_{\bd{q}}v_n(q).
\end{align}
The cohesive energy of the liquid phase $E_{coh}^L(n,s)$ is 
given, in terms of Haldane's pseudopotentials~\cite{haldane}
\begin{equation}\label{eq:haldanepp}
V^n_{2m+1}= \frac{1}{N_\Phi} \sum_{\bd{q}}v_n(q)L_{2m+1}(q^2\ell_B^2)e^{-q^2\ell_B^2/2},
\end{equation} 
by
\begin{equation} \label{eq:coh}
E^L_{coh}(n,s)=\frac{\bar{\nu}}{\pi}\sum_{m=0}^{\infty}c^s_{2m+1}V^n_{2m+1},
\end{equation}
where $c^s_{2m+1}$ are dimensionless coefficients that are subject to three sum rules that stem from charge neutrality, perfect screening and compressibility~\cite{gmp}. Usually the coefficients are calculated by Monte Carlo simulations using these sum rules as constraints~\cite{gmp}. However, they can also be calculated analytically assuming that $c^s_{2m+1}=0$ for $m\geq s+3$, together with the condition that the electrons repel each other at short distances, such that we can write $c^s_{2m+1}=-1$ for $m<s$~\cite{goerbig2002}. This procedure yields the coefficients given in Table~\ref{tab:cms}, which are used below to analytically calculate the energy of the Laughlin liquid~\footnote{This table was previously given in Ref.~\onlinecite{goerbig}, but we repeat it here because a few typos have been corrected.}.
\begin{table}[h!]
\begin{center}
\begin{tabular}{|c||c|c|c|c|c|c|c|}
\hline
&$c_1^s$&$c_3^s$&$c_5^s$&$c_7^s$&$c_8^s$&$c_{11}^s$&$c_{13}^s$\\
\hline
$s=1$&-1&17/32&1/16&-3/32&0&0&0\\
\hline
$s=2$&-1&-1&7/16&11/8&-13/16&0&0\\
\hline
$s=3$&-1&-1&-1&-25/32&79/16&-85/32&0\\
\hline
$s=4$&-1&-1&-1&-1&-29/8&47/4&-49/8\\
\hline
\end{tabular}
\caption{Coefficients $c^s_{2m+1}$, where $s$ represents the fractional filling $\bar{\nu}=1/(2s+1)$.}
\label{tab:cms}
\end{center}
\end{table}

For graphene, we can calculate these cohesive energies by substituting the effective potential $v_n^g(q)$ in the Haldane's pseudopotentials (\ref{eq:haldanepp}). To compute the total energy, we also have to substitute $v_n^g(q)$ in Eq.~\eqref{eq:utotal}. The analytic results for the total energy using these equations are shown in Table~\ref{tab:energies}, together with the results from exact-diagonalization calculations. (Details on the exact diagonalization method can be found in the Appendix  \ref{appED}). Note that the numerical result in Table~\ref{tab:energies} refers to the \emph{exact} ground state energy (per particle) of the Coulomb interaction, extrapolated to the thermodynamic limit (as $N_\Phi \to \infty$); in contrast, the analytic result is the \emph{variational} energy that is obtained when the trial state is assumed to be described by the Laughlin wave function. Although in general these two values can be different, we  can conclude that the theoretical and numerical calculations are in excellent agreement: all the values coincide within the range of uncertainty of the numerical calculations, except for $\bar{\nu}=1/3$ in the $n=2$ Landau level. However, this is to be expected if we compare the energies of the electron-liquid and electron-solid phases (see Section~\ref{sec:competition}), because for that filling the energy of the electron-solid is lower than that of the Laughlin liquid.

\begin{table}[]
\centering
\begin{tabular}{l|c|c|c|c}
 & \multicolumn{2}{c|}{$\bar{\nu}=1/3$}    & \multicolumn{2}{c}{$\bar{\nu}=1/5$}       \\ \hline
 &  analytic & numerical  	&                       analytic & numerical                     \\ \hline
$n=0$ & -0.409 &$-0.409 \pm 0.001$  & -0.327 & $-0.327 \pm 0.002$   \\
$n=1$ & -0.370 & $-0.369 \pm 0.001$ & -0.311 &$-0.311 \pm 0.002$   \\
$n=2$ & -0.265 & $-0.290 \pm 0.002$ & -0.273 &$-0.273 \pm 0.003$  \\
\end{tabular}

\begin{tabular}{c|c|c}
 & \multicolumn{2}{c}{$\bar{\nu}=1/7$}   \\ \hline
 &   analytic & numerical                     \\ \hline
$n=0$ & -0.280 &$-0.281 \pm 0.003$  \\
$n=1$ & -0.271&$-0.271 \pm 0.004$  \\
$n=2$ & -0.252 &$-0.251 \pm 0.005$
\end{tabular}
\caption{Comparison of our analytic results for the total ground state energies $U$ of the Laughlin liquid phase in graphene (in units of $e^2/\epsilon \ell_B$) with the exact ground state energies obtained numerically (see Appendix \ref{appED}).}
\label{tab:energies}
\end{table}

In the vicinity of the Laughlin states, we can use a first-order expansion to include the energy of the quasiparticle and quasihole excitations $\Delta_{\pm}$. The total cohesive energy of the quantum liquid then reads~\cite{goerbig}
\begin{align*}
E^{q-l}_{coh}(n,s,\bar{\nu}_{\pm})=E^L_{coh}(n,s)+[\pm\bar{\nu}(2s+1)-1]\Delta_{\pm}^n(s).
\end{align*}
The quasiparticle/quasihole excitation energies ($\Delta_+$/$\Delta_-$) in the $n$-th Landau level were derived using Murthy and Shankar's ``Hamiltonian theory"~\cite{shankar,murthyshankar}. They read
\begin{align*}
\Delta^n_+(s,p)=&\frac{1}{2}\int_{\bd{q}} v_n(q) \langle p|\bar{\rho}^p(-\bd{q})\bar{\rho}^p(\bd{q})|p\rangle \\
&- \int_{\bd{q}} v_n(q)\sum_{j'=0}^{p-1} |\langle p|\bar{\rho}^p(\bd{q})|j'\rangle|^2,\\
\Delta^n_-(s,p)=&-\frac{1}{2}\int_{\bd{q}} v_n(q) \langle p-1|\bar{\rho}^p(-\bd{q})\bar{\rho}^p(\bd{q})|p-1\rangle \\
&+ \int_{\bd{q}} v_n(q)\sum_{j'=0}^{p-1} |\langle p-1|\bar{\rho}^p(\bd{q})|j'\rangle|^2,
\end{align*}
where the matrix elements are given explicitly by~\cite{goerbig}
\begin{align*}
&\langle j|\bar{\rho}^p(\bd{q})|j'\rangle=\,\sqrt{\frac{j'!}{j!}}\left(\frac{-i(q_x-iq_y)\ell_B^*c}{\sqrt{2}}\right)^{j-j'}\\
&\times e^{-|q|^2\ell_B^{*2}c^2/4}\left[L_{j'}^{j-j'}\left(\frac{|q|^2\ell_B^{*2}c^2}{2}\right)\right.\\
&\left.-c^{2(1-j+j')}e^{-|q|^2\ell_B^{*2}/2c^2}L_{j'}^{j-j'}\left(\frac{|q|^2\ell_B^{*2}}{2c^2}\right)\right],
\end{align*}
with $\ell_B^*=\ell_B/\sqrt{1-c^2}$ and $c^2=2ps/(2p+1)$, with $s$ and $p$ integers. 

\begin{figure}[h!]
\centering
\includegraphics[width=\columnwidth]{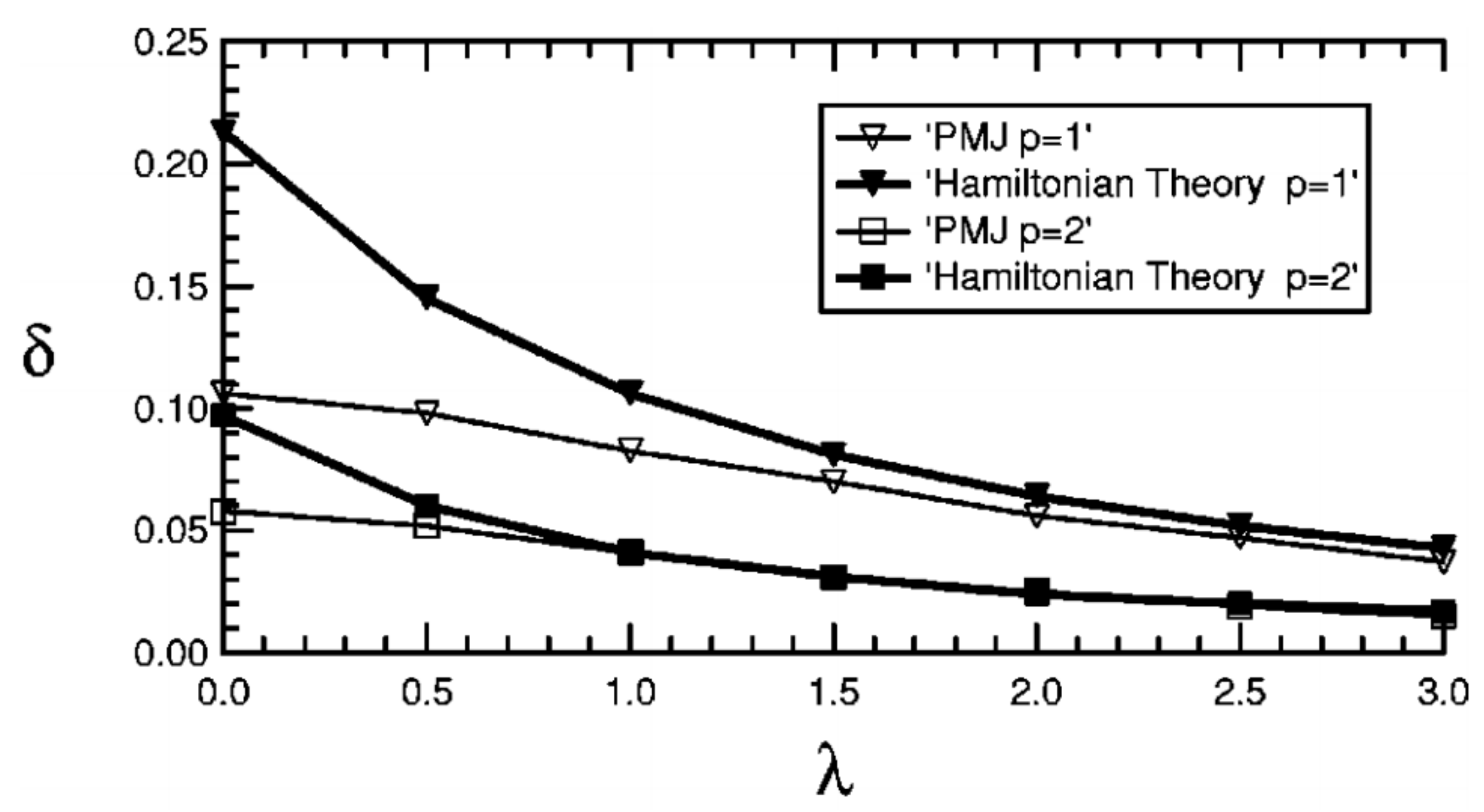}
\caption{Charge gaps $\delta$ [$\Delta$ in our notation] in the lowest Landau level calculated analytically using the Hamiltonian theory of Murthy and Shankar~\cite{shankar,murthyshankar} and numerically by Park et al.~\cite{park}. The $p=1$ case corresponds to the filling $\bar{\nu}=1/3$. The figure is reproduced from Ref.~[\onlinecite{shankar}].}
\label{fig:badresults}
\end{figure}

When we calculate the quasiparticle/quasihole energies for graphene by inserting the appropriate effective interaction $v_n^g(q)$, we do not obtain good agreement with the numerical results (see Table~\ref{tab:gaps}).  However, there are several possible complications that we must consider when interpreting this result. First, the finite-size extrapolation of numerical data is much less accurate for excited states, compared to the ground-state energy. Second, it should be noted that even for the conventional 2DES, there is a factor of 2 discrepancy between the predictions of the Hamiltonian theory and numerical calculations. In Fig.~\ref{fig:badresults}, this is illustrated by a comparison of the gaps $\Delta=\Delta_{+}+\Delta_{-}$ calculated analytically by the Hamiltonian theory of Murthy and Shankar~\cite{murthyshankar} and numerically by Park et al.~\cite{park} for a conventional 2DES in the lowest Landau level. The gaps are plotted as a function of the thickness parameter $\lambda$, which appears in the so-called Zhang-Das Sarma potential as~\cite{shankar} $v(q)=(2\pi e^2/\epsilon q) e^{-q \lambda}$. This phenomenological interaction mimics the ``softening" of the Coulomb interaction due to finite thickness of the 2DES. In our case, only $\lambda=0$ is relevant because graphene is atomically thin in the perpendicular direction. From this figure, we see that for $\lambda=0$ the gap calculated with the Hamiltonian theory is twice as large as the numerical one. Remarkably, the theory of Goerbig et al.~\cite{goerbig}, which uses the Hamiltonian theory to compute the quasiparticle and quasihole energies, reproduces the experimental phase diagram of GaAs very accurately~\cite{lewis}. We therefore adopt this approach as a first-order approximation to the competing solid and liquid phases in graphene Landau levels.

\begin{table}
\centering
\begin{tabular}{c|c|c|c}
LL & $\bar{\nu}$ & Numerical results & Analytic results\\
\hline
\multirow{3}{*}{$n=1$} & 1/3 & $0.116 \pm 0.002$ & 0.143\\
 & 1/5 & $0.020 \pm 0.002$ & 0.064\\
 & 1/7 & $0.007 \pm 0.001$ & 0.040\\
 \hline
\multirow{3}{*}{$n=2$} & 1/3 & $0.02 \pm 0.01$ & 0.114\\
 &  1/5 & $0.026 \pm 0.003$ & 0.067\\
 & 1/7 & $0.008 \pm 0.001$ & 0.048\\
\end{tabular}
\caption{Comparison of numerical and analytic results for the excitation gap $\Delta=\Delta_+ + \Delta_-$.}
\label{tab:gaps}
\end{table}

\section{Competition between phases}\label{sec:competition}

Now, we compare the energies of the solid phases with those of the liquid phase to determine the phase diagram in various Landau levels. Remember that in the lowest Landau level the energies in graphene are the same as in GaAs.

\begin{figure}
\centering
\includegraphics[width=\columnwidth]{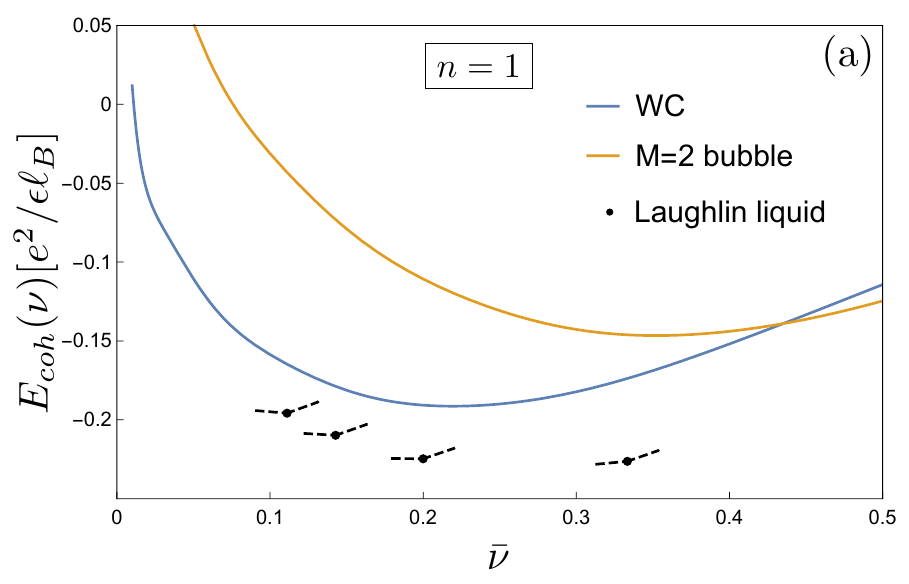}
\includegraphics[width=\columnwidth]{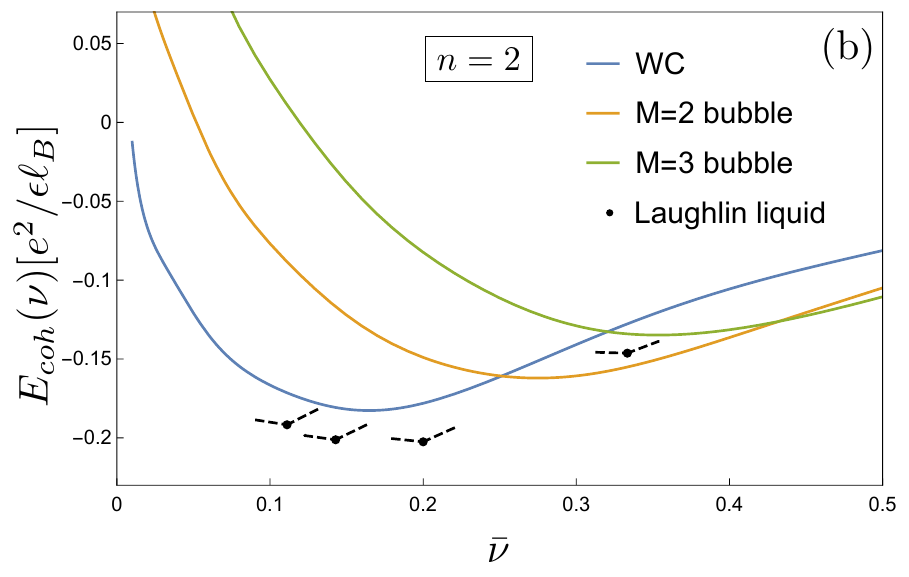}
\includegraphics[width=\columnwidth]{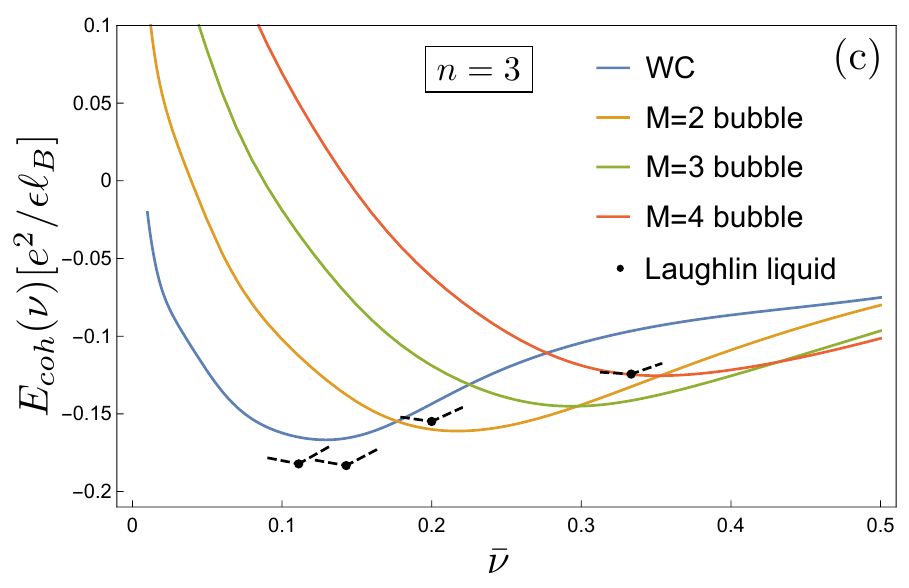}
\caption{(Color online) Energies of various bubble phases in the (a) $n=1$, (b) $n=2$ and (c) $n=3$ Landau level in graphene. The dashed lines denote the slope of a single quasiparticle or quasihole energy, computed by the Hamiltonian theory, relative to the ground state. Note that the plots show only the cohesive part of the total energy, therefore the liquid energies in this figure differ from Table \ref{tab:energies}.}
\label{fig:competition} 
\end{figure}

For the $n=1$ Landau level, the results are shown in Fig.~\ref{fig:competition}(a). The first thing to notice is that the Laughlin-liquid energies are always well below those for the Wigner crystal (blue solid line) or bubble phase (yellow solid line). Away from the filling factors $\bar{\nu}=1/(2s+1)$, the energies of the liquid phase are expected to be slightly higher than the Laughlin-liquid energy because of the incompressibility gap. However, because the slopes of the energies of the liquid phase, which are determined by the quasiparticle and quasihole energies, are so small, it is very unlikely that the liquid-phase energies will exceed the electron-solid energies. This case is actually quite similar to the lowest Landau level. The resemblance is understandable because the form factor in graphene is a combination of the $n$-th and the $(n-1)$-th form factor of GaAs, hence the lowest Landau level behavior may be dominating in this case.

For the $n=2$ Landau level shown in Fig.~\ref{fig:competition}(b), the FQHE is not expected to occur at $\bar{\nu}=1/3$, since the 2-electron-bubble phase is lower in energy. The other FQH states might be visible, however.

For the $n=3$ Landau level in Fig.~\ref{fig:competition}(c), the $\bar{\nu}=1/3$ and the $\bar{\nu}=1/5$ Laughlin states are always higher in energy than the bubble phases, thus they will not be visible. At higher filling factors, there are 2-electron-, 3-electron- and possibly 4-electron bubbles appearing and also coexisting, as it would be evident by performing a Maxwell construction. 

\subsection{The effect of impurities}
\begin{figure}
\centering
\includegraphics[width=\columnwidth]{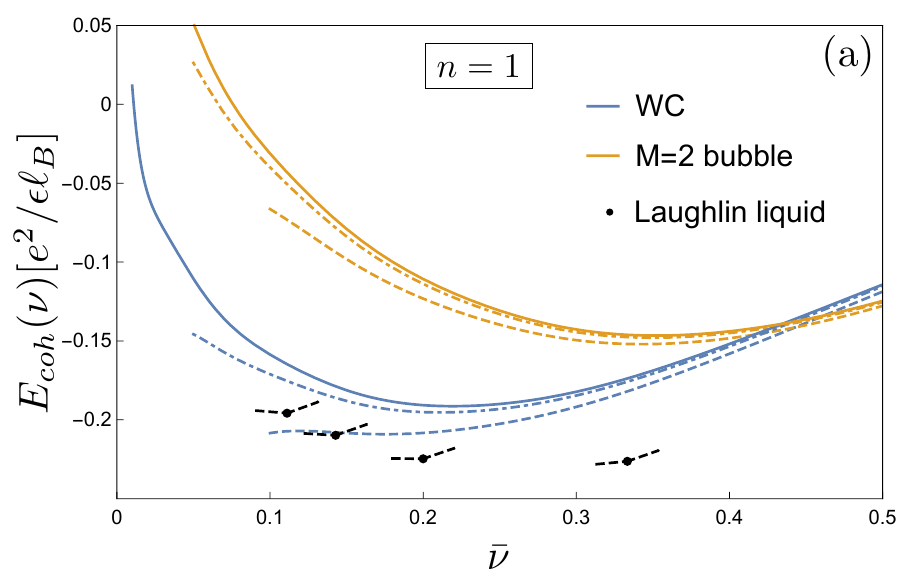}
\includegraphics[width=\columnwidth]{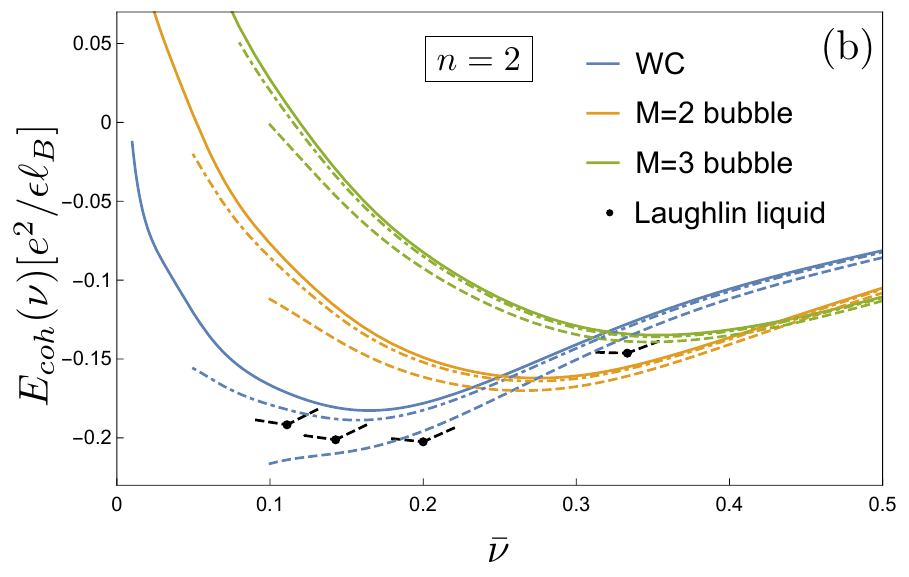}
\includegraphics[width=\columnwidth]{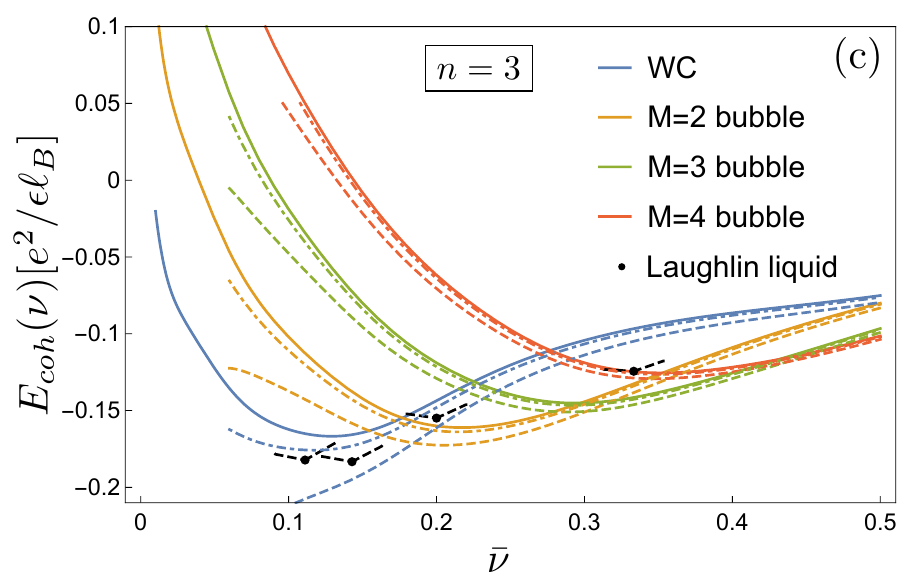}
\caption{(Color online) Energies of various bubble phases in the (a) $n=1$, (b) $n=2$  and (c) $n=3$ Landau level in graphene, where the dashed (dashed-dotted) lines represent the solid phases in an impurity potential with $E_{pin}=10^{-4}$ ($E_{pin}=2.5\cdot 10^{-5}$).}
\label{fig:impurities} 
\end{figure}

Until thus far, we have not considered impurities in the sample. These impurities can collectively pin the solids and lower their energy, while the incompressibility of the liquid phases makes them not very susceptible to impurities. Hence, we neglect the small change in the liquid energy due to disorder. To take the impurities into account, we model them by a Gaussian impurity potential with strength $V_0$ and correlation length $\xi$ in the weak-pinning limit. In this limit, the energy gained by individually following the impurity potential is small compared to the elastic energy it costs to deform the crystal. However, the energy gained by collectively following the impurity potential is large enough to overcome the elastic energy of the deformation. The energy density that describes this competition is given by~\cite{pin1,pin2}
\begin{align*}
\epsilon(L_0) = \frac{\mu\xi^2}{L_0^2}-V_0\frac{\sqrt{n_{el}}}{L_0},
\end{align*}
where $\mu$ is the elasticity of an $M$-electron bubble and is given by $\mu\approx 0.25M^2e^2n_M^{3/2}/\epsilon$, with $n_M=\bar{\nu}/2\pi M l_B^2$ the bubble density~\cite{goerbig}. Furthermore, $L_0$ is the Larkin length, which is the typical length scale on which the electron-solid is collectively pinned by impurities. Minimizing this energy with respect to $L_0$ yields the reduction of the solid-phase energy due to pinning by impurities, as derived in Ref.~\onlinecite{goerbig},
\begin{align*}
\delta E_{coh}^B(M,\bar{\nu})= -\frac{e^2}{\epsilon \ell_B}\frac{(2\pi)^{3/2}}{\sqrt{M}\bar{\nu}^{3/2}}E_{pin}^2,
\end{align*}
where the dimensionless pinning energy is defined as $E_{pin}=(V_0/\xi)/(e^2/\epsilon \ell_B^2)$ and $M$ is the number of electrons per bubble. We re-evaluate the phase diagrams with this energy, as shown in Fig.~\ref{fig:impurities}.

In the $n=1$ Landau level, the $\bar{\nu}=1/3$ and $\bar{\nu}=1/5$ liquid states are much lower in energy than the solid phases, even when taking the impurities into account. On the other hand, for a strong impurity potential the Wigner crystal phase can become the ground state for small filling factors. Reentrant behavior, where solid and liquid phases alternate with each other, is also conceivable if the impurity potential is very strong.

In the $n=2$ Landau level, the dashed lines in Fig.~\ref{fig:impurities}(b) show that if there are a lot of impurities in the sample, the FQH states at low fillings might be dominated by the Wigner crystal. The $\bar{\nu}=1/5$ state seems to be the most likely to be visible in this Landau level. For filling factors within the range $0.2-0.32$ there might also be a Wigner crystal coexisting with a 2-electron bubble. 
 
In the $n=3$ Landau level, the FQH states at lower fillings may only be seen in very clean samples, as indicated by the dashed lines in  Fig.~\ref{fig:impurities}c. Otherwise, the Wigner crystal will be the ground state for low filling and since the crystal is pinned by the impurities, the collective sliding mode is suppressed~\cite{goerbig}, which leads to a broadening of the plateau in the Hall resistance at integer filling.

\subsection{The effect of Landau-level mixing}
Besides sample impurities, there are some other factors that can have an influence on the energies of the liquid and solid phases. One of them is Landau-level mixing \cite{llmix1, llmix2, llmix3, llmix4, llmix5, llmix6, llmix7, llmix8, llmix9}. 
Thus far, we have assumed that there are no inter-Landau-level excitations, hence all dynamics was restricted to one single Landau level. However, if the Coulomb interaction energy becomes of the same order as the inter-Landau-level separation $\hbar v_F/\ell_B$, then these excitations can have a considerable probability, even if the Landau level is only partially filled. Note that the notion of Landau-level mixing should not be confused with the fact that the form factor in graphene is a sum of the usual form factors in the $n$-th and $(n-1)$-th Landau level. Although this may intuitively appear as a mixing of the two consecutive Landau levels, our previous discussion has neglected the explicit inter-Landau-level excitations.

To quantitatively characterize the effect of Landau level mixing, one introduces the mixing parameter $\kappa$, defined by 
\begin{align*}
\kappa = \frac{e^2/\epsilon \ell_B}{\hbar v_F/\ell_B} = \frac{e^2}{\epsilon\hbar v_F}.
\end{align*}
The only parameters that determine the value of $\kappa$ are the Fermi velocity $v_F$ and the dielectric constant $\epsilon$. These are material properties, which depend on the substrate. For free-standing graphene $\kappa\approx 2.2$, while on substrates such as SiO$_2$ or BN, it takes slightly lower values $\kappa\approx 0.9$ and $\kappa\approx 0.5-0.8$, respectively~\cite{llmix6}. 
In Ref.~[\onlinecite{llmix6}], the effect of Landau level mixing in graphene on the Laughlin-liquid states in the $\bar{\nu}=1/3$ case is investigated in the $n=0$ and $n=1$ Landau level. This effect turns out to be negligible for $\kappa\lesssim 2$ in the lowest Landau level and for $\kappa\lesssim 1$ in the $n=1$ Landau level. Hence, for not too large values of $\kappa$ most states will be unaffected. However, the Landau level mixing is expected to get stronger for higher values of $n$. Additionally, the corrections to the renormalized Coulomb interaction due to Landau level mixing include not only two-body terms, but also three body (and higher-order) terms, which break particle-hole symmetry. Such terms could affect, for example, the stability of $\bar{\nu}=1/3$ state relative to $\bar{\nu}=2/3$. The full treatment of Landau-level mixing effects, however, is beyond the scope of present work.

On the level of the Hartree-Fock approximation, which we expect to hold for the solid phases, we note that Ref.~[\onlinecite{zhangmix}] has investigated the validity of the single-Landau-level approximation. In turns out that there are no qualitative changes in the phase diagram, e.g., in the $n=2$ Landau level there are still phase transitions from Wigner crystal to 2-electron bubble and then to 3-electron bubble, but these transitions might occur at slightly different filling factors. Furthermore, the cohesive energies stay in the same range as they were (within about $10\%$). As a last remark, we note that the inter-Landau-level spacing becomes smaller for higher Landau levels, since it scales like $\sqrt{n+1}-\sqrt{n}$, which goes to zero for increasing $n$, and one might argue that Landau-level mixing then becomes dominant. However, in Ref.~[\onlinecite{zhangmix}] it was also shown that this argument is too simplistic, and it turns out that the single-Landau-level approximation stays in fact applicable, even for large $n$.

\section{Conclusions}\label{sec:conclusion}
We calculated the energies of various electronic phases in partially-filled Landau levels in graphene. For the liquid phase, we compared our analytic calculations for the electron-liquid energies to the exact values obtained by numerical diagonalization and finite-size scaling. We found excellent agreement between the two approaches. For the lowest Landau level ($n=0$), graphene behaves in the same way as GaAs, which exhibits no charge-density waves but only electron-liquid phases. Thereby, the fractional quantum Hall effect can be observed at filling factors $\bar{\nu}=1/(2s+1)$, for integer $s$. For the $n=1$ Landau level in graphene, we have seen that there are still some characteristics of the lowest Landau level present, as a consequence of the Landau level averaging in the form factor. We have shown that the liquid phases are lower in energy than the solid phases for all filling fractions. However, impurities in the sample can lower the energy of the electron-solid phases, especially at low filling. Indeed, if the impurity potential is sufficiently strong, the $\bar{\nu}=1/7$ and $\bar{\nu}=1/9$ fractional quantum Hall states can be washed out. The plateau in the resistivity will then be broadened. In higher Landau levels, the electron-solid phases become more pronounced. In the $n=2$ Landau level, for instance, the $\bar{\nu}=1/3$ fractional quantum Hall state is dominated by the 2-electron-bubble phase, even without impurities. As the Landau level increases, more bubble phases emerge and there might also be phase coexistence between different electron-solid phases. From the phase diagrams that we constructed, one can determine the filling factors for which there might be electron bubbles and in future experiments one can attempt to measure these bubble phases to verify existing predictions that were done using numerical calculations on the local density of states~\cite{poplavskyy}.

In this work, we have assumed that graphene's SU(4) symmetry~\cite{fngraphene} is completely broken, and the fractional states mentioned above correspond to partial fillings of a sublevel with the given spin- and valley quantum number. Under SU(4) symmetry, integer quantum Hall states in graphene occur at experimental filling factors $\nu_0=\pm 2 (2k+1)$, $k=0,1,2,3,\ldots$. Thus, 
our Landau level $n=0$ corresponds to the range of experimental filling factors $-2 \leq \nu_0 \leq 2$, Landau level $n=1$ on the electron side is  $2 < \nu_0 \leq  6$, Landau level $n=2$ is  $6 < \nu_0 \leq 10$, etc. In experiment, our results suggest that first possible signatures of electron solid phases occur at fillings around or above $\nu_0 + 0.3$ for $\nu_0 \in \{ 6, 7, 8, 9\}$. Moreover, electron-solid phases are expected to become more prominent at fillings above $\nu_0 + 0.2$, $\nu_0 \geq 10$.

Although charge-density waves have been observed in graphene, the underlying mechanisms are very different to the one described in this work. For example, charge-density waves were observed in twisted graphene layers~\cite{andrei}. By twisting the two layers of bilayer graphene, the van Hove singularities are lowered and charge-density waves become experimentally observable. In recent Hofstadter butterfly experiments in graphene superlattices~\cite{dean}, there were also signs of charge-density waves. However, in monolayer graphene, there have been no experimental signs thus far of a charge-density wave that arises due to electronic interactions in a quantum Hall system, which is the system that we have described here. It would be interesting to see whether future experiments in higher Landau levels can verify our theoretical predictions of the various electron phases in graphene, just like they did for GaAs.

\appendix

\section{Exact diagonalization results}\label{appED}

\begin{figure*}[ttt]
  \begin{minipage}[l]{\linewidth}
\includegraphics[width=0.95\linewidth]{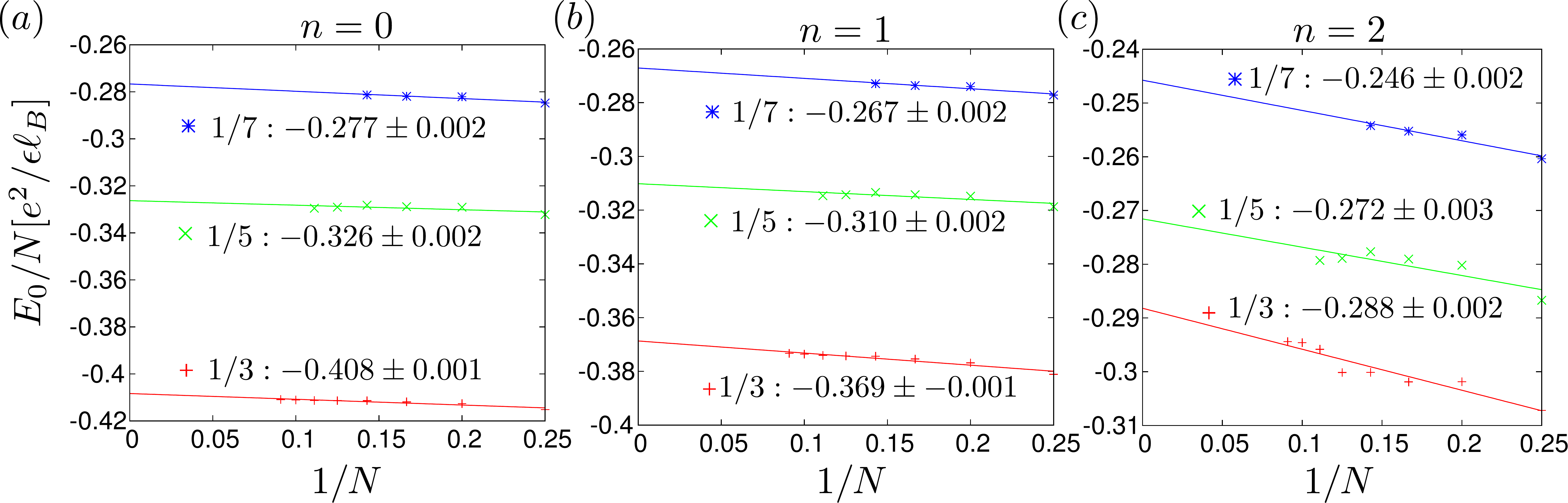}
  \end{minipage}
\caption{(Color online) Finite-size scaling of ground-state energy per particle in graphene Landau levels (a) $n=0$, (b) $n= 1$, and (c) $n= 2$.  In each Landau level, we compute the ground-state energy at filling fraction $\nu=1/3$, $1/5$, and $1/7$. The unit cell of the 2DES is a square with area equal to $2\pi\ell_B^2 N/\nu$, where $N$ is the number of electrons. The finite-size extrapolations are fairly accurate, with the exception of the $n=2$ Landau level, where we expect the ground state to break translational symmetry.}
\label{fig:edgs}
\end{figure*}

For the purposes of numerical finite-size calculations, we must choose an appropriate geometry (boundary condition). Since the FQHE occurs in an effectively continuum system of electrons, various
choices are possible and have been used in the literature: finite disk, sphere, and periodic boundary conditions, either along one direction (cylinder geometry) or both $x$ and $y$-directions (torus). 

\begin{figure}[htb]
\includegraphics[width=0.7\linewidth]{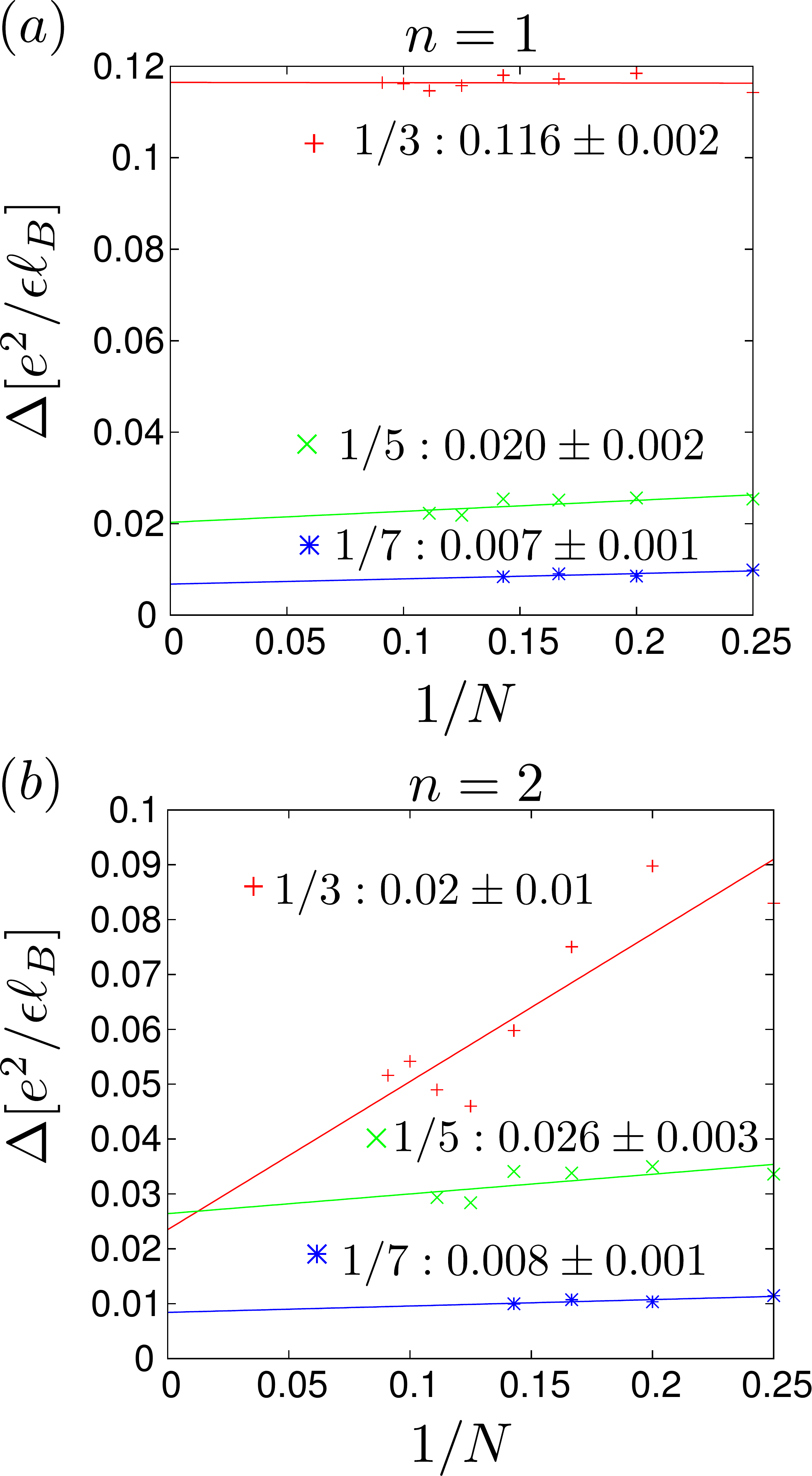}
\caption{(Color online) Charge gaps in the (a) $n=1$ and (b) $n=2$ graphene Landau levels for filling fractions $1/3$, $1/5$, and $1/7$. The unit cell of the 2DES is a square with area equal to $2\pi\ell_B^2 N/\nu$, where $N$ is the number of electrons. Extrapolations are accurate in the $n=1$ Landau level, but less so in the $n=2$ Landau level, where the ground state likely breaks translation symmetry.}
\label{fig:edgap}
\end{figure}

Our goal is to compute the ground-state energy and excitation gap, which are bulk properties. Therefore, it is convenient to choose a closed system in order to eliminate edge effects. This rules out finite disk and cylinder geometry. In many cases, sphere geometry would be the simplest choice because the ground state of incompressible  liquids is always unique (while on a torus, it has a degeneracy of at least $q$ for filling fraction $\nu=p/q$). Unfortunately, due to the more complicated form of graphene form-factors in Eq.~(\ref{eq:graphenepp}), we expect that the corresponding Haldane pseudopotentials on a finite sphere must be computed numerically. More importantly, because we consider a competition between electron-liquid and -solid phases, it is essential that they are treated on the same footing. This is difficult to achieve in the spherical geometry, which intrinsically favors liquid phases over crystalline ones. Hence, for our purposes the torus geometry remains as the most natural choice for finite-size calculations. 

The description of FQHE in the torus geometry is introduced and explained in detail in Refs.~[\onlinecite{haldane_book, qh_book}]. The 2DES is placed on a surface $L_x \times L_y$ with periodic boundary conditions along both $x$ and $y$-directions. Because of the magnetic field, the system is invariant under \emph{magnetic} translations, which are ordinary translations followed by a gauge transformation. Haldane has shown \cite{haldane_book} how to construct an eigenbasis in the many-body Hilbert space that is invariant under such transformations. This also results in a conserved quantum number -- the ``pseudomomentum" -- which labels all many-body states. The final Hamiltonian is similar to Eq.~\eqref{eq:ham} except that the density operators must also be compatible with the magnetic translations, and the Coulomb potential must be periodic in the unit cell. This periodicity implies that in order to calculate the energy of the 2DES, we must include a correction due to the interactions of an electron with its periodic images, known as the Madelung term \cite{qh_book}.

Using iterative (Lanczos) diagonalization, in Fig.~\ref{fig:edgs} we compute the ground-state energy of a finite 2DES containing $N$ electrons in the $n=0, 1, 2$ Landau levels of graphene. In all the calculations, the unit cell of the 2DES is a square with area equal to $2\pi\ell_B^2 N/\nu$. In each of the Landau levels, we obtain the ground state for three filling factors $\nu=1/3, 1/5, 1/7$. By performing a simple $1/N$ fit, we are able to obtain accurate estimates of the energy per electron in the thermodynamic limit. These values agree with the previous data in the literature (where available) and with our analytic results in Sec. \ref{sec:liquid}. The extrapolation is less trustworthy in the $n=2$ Landau level; here we expect the ground state to break translation symmetry and therefore its energy should be very sensitive to the shape of the unit cell. By imposing the square unit cell, we have likely introduced another finite-size effect; this could possibly be minimized by looking for a global energy minimum between all possible shapes of the unit cell for each given $N$, but we have not performed such an exhaustive search.

Additionally, by diagonalizing the system in sectors of the Hilbert space that contain $N$ electrons in an area of size $2\pi\ell_B^2 (N/\nu \pm 1)$, we can obtain the energy of the system with one added quasiparticle (-) or quasihole (+).  In these cases, in order to get the total energy, we must carefully correct for the fact that a quasiparticle or quasihole also interacts with own images due to periodic boundary conditions. Assuming the excitation is a point object (i.e., much smaller in area than $2\pi\ell_B^2 N/\nu$, which is justified for the Laughlin states), this correction can be computed in the way explained in Ref.~[\onlinecite{xinwan}].  Accounting for this, in Fig.~\ref{fig:edgap} we plot the ``charge gaps" defined as $\Delta = E_{N, 2\pi\ell_B^2 (N/\nu+1)} + E_{N, 2\pi\ell_B^2 (N/\nu-1)} -2 E_{N, 2\pi\ell_B^2 N/\nu}$ in the $n=1$ and $n=2$ Landau levels of graphene. Typically, the finite-size effects in the scaling of gaps are worse than for the ground-state energy. We see that our extrapolations are still rather accurate in the $n=1$ Landau level, but become significantly worse in the $n=2$ Landau level. Similar to the energy calculation, if the ground state in the $n=2$ Landau level is not a liquid, there is no reason to assume that the optimal shape of the unit cell should be a square. This may be partly responsible for the poorer scaling of the data in this case. The fact that the system displays such sensitivity to the geometry is a strong indication that the ground state is not a liquid.

\acknowledgments 

We acknowledge fruitful discussions with M. Goerbig, A. Quelle, G. van Miert, V. Juri$\check{\text{c}}$i\'{c} and Cory Dean. ZP acknowledges support by DOE grant DE-SC0002140. This work is part of the DITP
consortium, a program of the Netherlands Organisation
for Scientific Research (NWO) that is funded by the Dutch
Ministry of Education, Culture and Science (OCW).

\end{document}